\documentclass[12pt, a4paper]{iopart}
\usepackage{amssymb}
\usepackage{hyperref}
\usepackage{xurl}
\usepackage{graphicx}
\usepackage{geometry}

\begin{document}
\title{Energy recovery in filament-regime plasma wakefield acceleration of positron beams}

\author{Max Varverakis$^{1}$, Robert Holtzapple$^1$, Severin Diederichs$^2$, Carl Schroeder$^3$ and Spencer Gessner$^4$}

\address{$^1$ Department of Physics, California Polytechnic State University, 1 Grand Ave, San Luis Obispo, 93407, CA, USA}
\ead{mvarvera@calpoly.edu}

\address{$^2$ Deutsches Elektronen-Synchroton {DESY}, Notkestraße 85, 22607, Hamburg, Germany}

\address{$^3$ Lawrence Berkeley National Laboratory, 1 Cyclotron Road, Berkeley, 94720, CA, USA}

\address{$^4$ {SLAC} National Accelerator Laboratory, 2575 Sand Hill Road, Menlo Park, 94025, CA, USA}
\ead{sgess@slac.stanford.edu}

\begin{abstract}
    Plasma wakefield acceleration using an electron filament offers stable, high-gradient, high-quality acceleration of positron beams analogous to the acceleration of electrons in the blowout regime. However, low energy-transfer efficiency is currently a limiting factor for future collider applications. We explore the addition of a secondary electron bunch in the electron filament plasma wakefield acceleration scheme to recover additional energy from the wake. Particle-in-cell simulations using HiPACE++ are used to demonstrate various energy recovery schemes. In addition to confirming the energy efficiency gains with a recovery electron beam, we also develop energy recovery schemes in the context of future plasma colliders.
\end{abstract}

\vspace{2pc}
\noindent{\it Keywords}: Positron, Collider, Plasma 

\submitto{\JPCS}

\ioptwocol

\section{Introduction}
Plasma-based accelerators offer extremely high accelerating gradients, which may enable a compact future linear collider at the TeV scale~\cite{Rosenzweig1991, Chen2020, Schroeder2010, Aldi2013, Schroeder2023}.
In recent decades, progress in plasma wakefield acceleration (PWFA)~\cite{Blumenfeld2007, Litos2014} has addressed many challenges on the path to a future plasma collider. For electron-positron colliders, the ability to accelerate both electrons and positrons is imperative.
In the non-linear blowout regime, high-efficiency and high-quality PWFA can be achieved for electrons~\cite{Tzoufras2008} and has been demonstrated experimentally~\cite{Lindstrom2021}.
However, high-efficiency and high-quality positron acceleration remains a challenge in the non-linear regime~\cite{Hogan2003,Blue2003,Muggli2008,Corde2015, Doche2017}.

In the non-linear blowout case, the region of the wakefield that is both accelerating and focusing for positron bunches is relatively small. Recently, new ideas have been proposed to elongate the region of high plasma electron density at the back of the bubble, enabling better loading and acceleration of trailing positron bunches~\cite{Diederichs2019,Wang2021,Zhou2022}. Of these approaches, the plasma column case has been studied in the most detail~\cite{Diederichs2019}. 
The wake formed in the plasma column regime not only focuses and accelerates positrons, but it also preserves beam quality~\cite{Diederichs2020} and stabilizes under misalignment~\cite{Diederichs2022}. 


The plasma column regime is a promising candidate for positron acceleration in a future plasma collider, but the energy-transfer efficiency from the drive beam to witness beam is limited to roughly 5\% due to beam loading effects~\cite{Diederichs2020}. This indicates that most of the energy in the wake is not used to accelerate the witness beam. In the ongoing efforts towards minimizing the carbon footprint of particle accelerators, energy efficiency remains a crucial factor in determining the feasibility of a future plasma collider~\cite{Roser2023, Roser2022}. Therefore, efficiency enhancements for positron acceleration in the plasma column regime must be realized before this scheme can be utilized as part of a future plasma collider.

Other PWFA schemes that elongate the electron filament at the back of the bubble are currently under investigation, including the elongated bubble regime~\cite{Wang2021} and the uniform non-linear regime~\cite{Zhou2022}. In the uniform non-linear regime, the placement of a positron beam just behind the bubble of a blowout wake elongates the on-axis plasma electrons into an electron-dense filament capable of focusing, accelerating, and preserving the positron beam. In an optimized drive-witness configuration, efficiencies of up to 35\% have been demonstrated in simulations, which is comparable to the efficiency required for future plasma colliders~\cite{Chen2020}. Once again, minimizing energy losses is pertinent for a sustainable future collider, and maximizing energy efficiency remains an important step in the realization of a future plasma collider. 


Energy recovery from laser-driven plasma wakefields using trailing laser pulses has been proposed~\cite{Schroeder2016}. Extending this concept to beam-driven plasma wakefields, we propose the addition of a second electron beam to absorb additional energy from the wakefield. We use the 3D quasi-static particle-in-cell (PIC) code {HiPACE}$++$~\cite{HiPACE++} to show that the presence of an electron recovery beam can result in a significant increase in energy extraction from the wake.
Additionally, we develop future collider concepts with the energy recovery electron filament schemes. 


\section{High efficiency in the linear regime}

Before addressing energy recovery in the plasma column and uniform non-linear regimes, it is instructive to review energy recovery in the linear regime~\cite{Hue2021}. Perturbatively solving for the plasma response to an electron drive beam, it is possible to tailor a witness beam profile such that the fields are nearly zero behind the trailing bunch~\cite{Katsouleas1987}.
Given a drive beam density $\rho(\xi)$ and trailing bunch head at $\xi=\xi_0$, the requirement of a constant accelerating field $E_a$ across the witness beam is described by
\begin{equation}
    E_a = E_0\cos{k_p\xi}-4\pi\int_{\xi_0}^\xi d\xi'\rho(\xi')\cos k_p(\xi-\xi'),
\end{equation}
where the relevant variables are defined in \ref{appendix}.
As a result, for any positive integer $n$, loading an identical electron witness beam $(2n-1)\pi$ plasma skin depths behind the drive beam results in near-100\% energy-transfer efficiency from the wake to the witness beam. A positron beam can be loaded in a similar fashion, with $2n\pi$ plasma skin depth separation leading to ideal energy transfer.

\begin{figure*}
    \includegraphics[width = \linewidth]{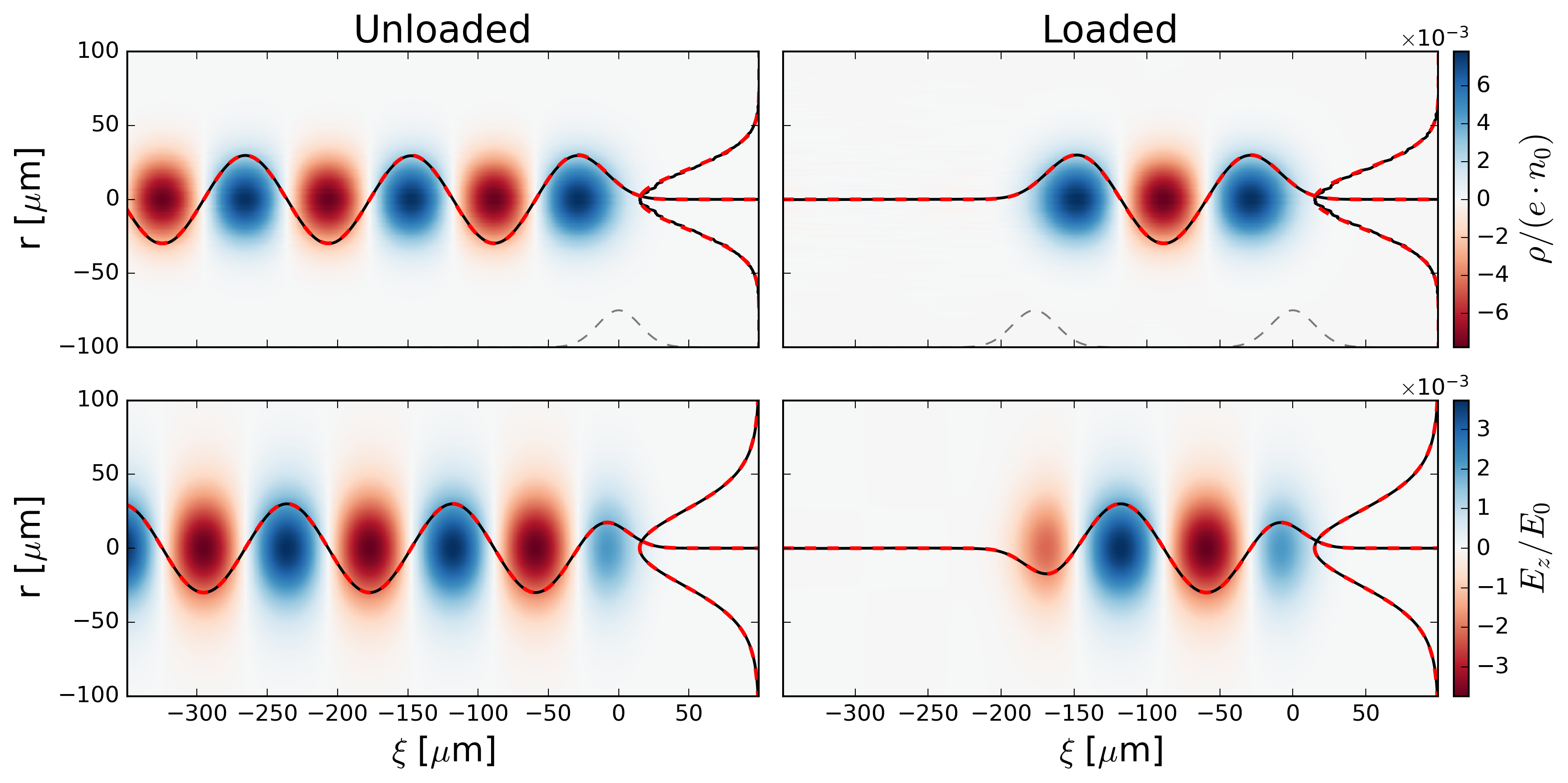}
    \caption{Normalized plasma charge density (top row) and accelerating field (bottom row) for a loaded (left column) and unloaded (right column) wake in the linear regime in the $x\xi$-plane. The on-axis fields from HiPACE$++$ are outlined in black and are in agreement with the analytical solutions indicated by red dashed lines. Gray dashed lines give the longitudinal current density profiles for the beam(s). The witness beam is loaded $-3\pi\,k_p^{-1}$ behind the drive beam. The simulation domain contains 1 plasma electron macro particle per cell and is $(-150, 150)\times(-150, 150)\times(-350, 100)~\mu$m${}^3$ in $x\,\times\,y\,\times\,\xi$. The resolution is $0.59\times0.59\times0.89~\mu$m${}^3$.}
    \label{fig:linear}
\end{figure*}

In HiPACE$++$, we simulate PWFA in the linear regime with two identical bi-Gaussian electron beams separated by $3\pi k_p^{-1}$ in $\xi$.
The simulation consisted of a plasma electron density of $n_0 = 8\times10^{16}$ cm$^{-3}$, $N_b = 4\times10^7$ macro particles per beam, and beam size $\sigma_{x,y} = 20$ $\mu$m and $\sigma_z = 15$ $\mu$m. Beams are injected at an energy of 10 GeV and a peak density of $4.23\times10^{14}$ cm${}^{-3}$.

Both simulation and numerical analysis show a full depletion of the wakefield at 100\% efficiency when properly loaded, as illustrated in Figure~\ref{fig:linear}. Unfortunately, the linear regime is not suitable for collider applications, because the low-emittance, high-charge, and high-energy beams are extremely dense and will drive non-linear wakes~\cite{Hue2021,Cao2023}.

\section{Energy recovery in the Plasma Column regime}
As thoroughly studied in~\cite{Diederichs2019, Diederichs2020, Diederichs2022, Diederichs2023}, a plasma column that is radially smaller than the blowout radius causes a spread in the plasma electron trajectories near the boundary of the ion bubble. The plasma wake forms a high-density plasma electron filament that is elongated on-axis. The filament region not only contains the necessary focusing and accelerating fields for positrons, but it also preserves beam quality and provides stability. 
To study energy recovery in the plasma column regime, we first simulate positron PWFA in the plasma column without energy recovery modifications to establish a baseline efficiency.
Next, we explore increasing the energy efficiency by placing an electron recovery bunch either in front of or behind the positron witness bunch. There are focusing and accelerating fields for a second electron bunch in both regions of the wake.
We note that the parameter space for energy recovery is large, with bunch charge, bunch length, and transverse emittance all free parameters. 
The goal of this study is to maximize energy extraction from the wake with a triple-beam configuration, without addressing all beam quality considerations.

Helium plasma is simulated with 400 electron and 16 ion macro particles per cell with a column radius of $2.5~k_p^{-1}$ and background density $n_0 = 10^{17}$ cm${}^{-3}$. 
A plasma temperature of $k_BT = 15$ eV is used, where $k_B$ is the Boltzmann constant. All three simulations use a -3.38 nC bi-Gaussian electron drive beam with $10^6$ macro particles, an initial energy of 5.11 GeV, and a $\beta$-matched normalized transverse emittance of $\epsilon_{x,y} = 2.96~\mu$m~rad. The drive beam is centered at $\xi=0$ with $\sigma_{x,y} = 0.05~k_p^{-1}$ and $\sigma_z=1.41~k_p^{-1}$. The positron witness and electron recovery beams are simulated with $1.25\times10^6$ and $10^6$ macro particles, respectively, and at energies approximately equal to 1 GeV. Both trailing beams are radially Gaussian with $\sigma_{x,y} = 0.029~k_p^{-1}$ and $\epsilon_{x,y} = 0.45~\mu$m rad for the positron beam and $\sigma_{x,y} = 0.05~k_p^{-1}$ and $\epsilon_{x,y} = 1.33~\mu$m rad for the recovery electron beam. Additional beam properties are listed in Table~\ref{tab:columnBeams}.



\begin{table}
    \caption{Trailing beam parameters for plasma column simulations. Subscripts $p$ and $r$ correspond to the positron and electron (recovery) beam, respectively.}
    \label{tab:columnBeams}
    \lineup
    \centering
    \begin{tabular}{cccc}
        \br
        Simulation & (a) & (b) & (c) \\
        \mr
        $k_p\xi_{p,\,\textrm{\footnotesize head}}$ & $-10.5$ & $-10.5$ & $-12.9$ \\
        $k_p\xi_{r,\,\textrm{\footnotesize head}}$ & -- & $-20.0$ & $\0\0\0\-7.6$ \\
        $Q_p$ [pC] & 182 & 181 & \064 \\
        $Q_r$ [pC] & -- & $\-517$ & $\-707$ \\
        $\eta$ [\%] & \03.8 & \012.0 & \027.4 \\
        \br
    \end{tabular}
\end{table}

\begin{figure*}[htbp]
    \includegraphics[width = \linewidth]{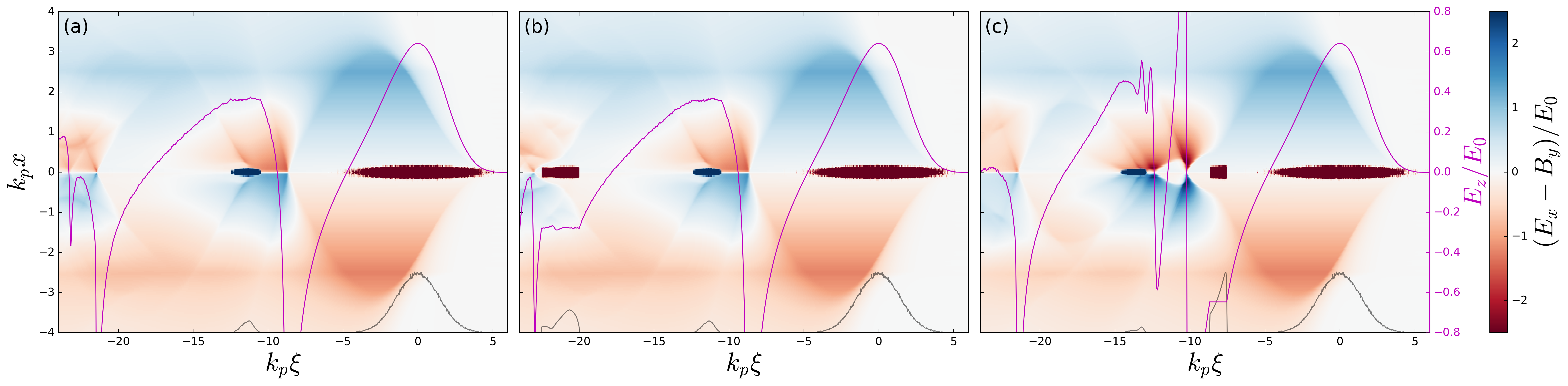}
    \caption{Normalized focusing wakefield (in $\xi$-$x$) and on-axis normalized accelerating field (magenta curve) in the plasma column filament regime for (a) no energy recovery, (b) recovery behind the positron beam, and (c) recovery in front of the positron beam. Longitudinal currents of electron beams are plotted in red and positron beams in blue. On-axis beam current density profiles are displayed on the $\xi$-axis in gray. Simulations were executed in a $(-16, 16)\times(-16, 16)\times(-24, 6)~k^{-3}_p$ domain in $x\,\times\,y\,\times\,\xi$. The corresponding mesh resolution is $0.031\times0.031\times0.029~k^{-3}_p$. }\label{fig:filamentWF}
\end{figure*}

Compared to Figure~\ref{fig:linear}, which shows near-100\% efficiency in the linear regime, Figure~\ref{fig:filamentWF} illustrates an imperfect energy transfer due to the fact that the wakefield persists behind the trailing beams. Nevertheless, the energy-transfer efficiency increases with the addition of a recovery bunch. Based on these simulations,  we find a baseline efficiency of $\eta = 3.8\%$. The additional electron recovery beams placed behind and in front of the positron beam resulted in efficiencies of $\eta = 12.0\%$ and $\eta = 27.4\%$, respectively. See \ref{appendix} for more detail on efficiency calculations. 

An iterative search was used to determine the electron recovery beam parameters that maximize the efficiency of the three-bunch configuration. The trailing bunch profiles were optimized to flatten the average accelerating fields with the {SALAME} algorithm as part of the HiPACE++ code~\cite{Diederichs2020}. Figure~\ref{fig:filamentWF}(c) shows large variation of the on-axis $E_z$ field for the trailing recovery bunch, but this variation is minimized by averaging $E_z$ transversely across the bunch. According to our simulations, transverse witness beam properties did not significantly affect energy-transfer efficiency for the initial particle injection. However, efficiency losses are expected to occur when propagating the beams with inappropriate transverse characteristics, such as when using a positron bunch transversely larger than the electron filament.

We note that our recovery beam parameter search was non-exhaustive, but initial results from our simulations indicate that the efficiency of the three-bunch scheme is unlikely to improve far beyond what we have demonstrated here. 

\section{Energy recovery in the Uniform Nonlinear regime}
Generating an electron filament in uniform plasma involves a slightly different procedure than in the plasma column case. 
In a traditional blowout wake, a region of highly dense plasma electrons appears behind the ion bubble, but is longitudinally short. 
Similar to the elongated bubble scheme~\cite{Wang2021}, a method was studied in~\cite{Zhou2022} in which the presence of a high-intensity positron beam causes an electron filament to form in the region behind the bubble of the blowout wake. The forces induced by the positron beam bring sheath electrons back on-axis to produce the desired electron filament. 

As before, there are three situations that we explore. First, we simulate this scheme without the use of an energy recovery beam to obtain a baseline in energy-transfer efficiency. Next, we add an electron recovery beam behind the positron beam. Finally, we simulate an electron recovery beam ahead of the positron beam.

\begin{figure*}
    \includegraphics[width = \linewidth]{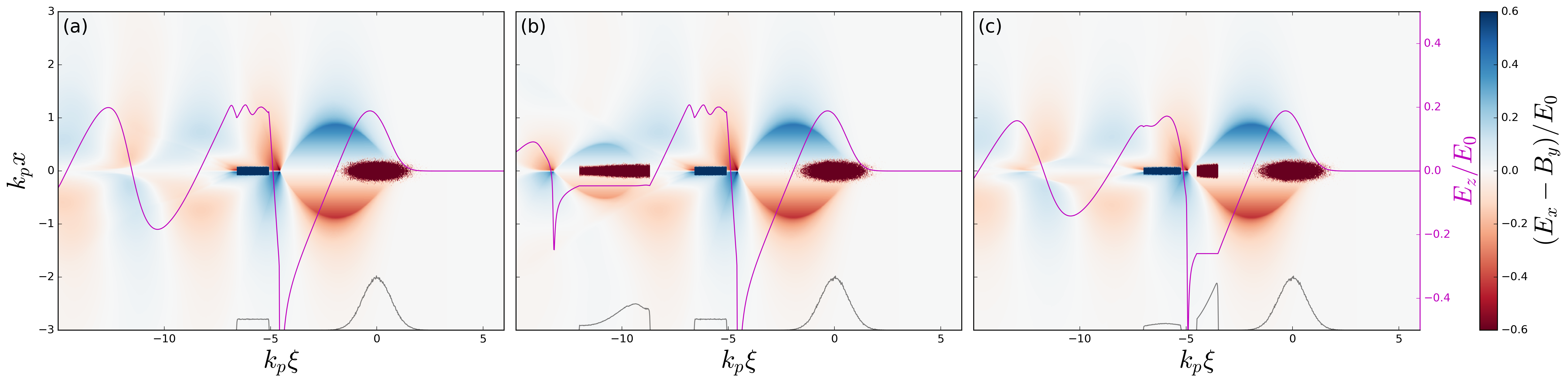}
    \caption{Normalized focusing wakefield (in $\xi$-$x$) and on-axis normalized accelerating field (magenta curve) in the uniform plasma filament regime for (a) no energy recovery, (b) recovery behind the positron beam, and (c) recovery in front of the positron beam. Longitudinal currents of electron beams are plotted in red and positron beams in blue. On-axis beam current density profiles are displayed on the $\xi$-axis in gray. We use a simulation window of $(-6, 6)\times(-6, 6)\times(-15, 6)~k^{-3}_p$ and a resolution of $0.012\times0.012\times0.021~k^{-3}_p$ in $x\times y\times\xi$. }\label{fig:uniformWF}
\end{figure*}

In the cases presented in Figure~\ref{fig:uniformWF}, we use a helium plasma with background density $n_0 = 7.8\times10^{15}$cm${}^{-3}$ and with 25 macro particles per cell for both plasma electrons and ions. All beams were simulated with $10^6$ macro particles and at 2.5 GeV. The beam parameters were chosen to match the parameters used in Ref.~\cite{Zhou2022}. The drive beam follows a bi-Gaussian profile with $\sigma_{x,y} = 5~\mu$m and $\sigma_z = 40~\mu$m. Both the positron beam and recovery beam are radially Gaussian with $\sigma_{x,y} = 2~\mu$m and $\sigma_{x,y} = 3~\mu$m, respectively. The total charge of the drive beam is $Q_d = -534$ pC. The drive beam is simulated with a normalized transverse emittance of 6 $\mu$m rad, the positron beam with 2.5 $\mu$m rad, and the recovery beam with 7.5 $\mu$m rad. See Table~\ref{tab:uniformBeams} for trailing beam placements and the corresponding energy-transfer efficiency.

\begin{table}
    \caption{Trailing beam parameters for uniform plasma simulations. Subscripts $p$ and $r$ correspond to the positron and electron (recovery) beam, respectively.}
    \label{tab:uniformBeams}
    \lineup
    \centering
    \begin{tabular}{cccc}
        \br 
        Simulation & (a) & (b) & (c) \\
        \mr 
        $k_p\xi_{p,\,\textrm{\footnotesize head}}$ & $-5.1$ & $-5.1$ & $-5.3$ \\
        $k_p\xi_{r,\,\textrm{\footnotesize head}}$ & -- & $-8.7$ & $-3.5$ \\
        $Q_p$ [pC] & 102 & 102 & \063 \\
        $Q_r$ [pC] & -- & $\-310$ & $\-177$ \\
        $\eta$ [\%] & \025.9 & \045.0 & \073.5 \\
        \br 
    \end{tabular}
\end{table}

As before, the presence of a wakefield behind the trailing beam in Figure~\ref{fig:uniformWF} indicates that the wake energy is not fully transferred to the witness beams. Nevertheless, the addition of the recovery bunch significantly increases energy extraction from the wake. The base case with no energy recovery resulted in an energy-transfer efficiency of $\eta = 25.9\%$, which is consistent with the findings in~\cite{Zhou2022}. Energy-transfer efficiency increases to $\eta = 45.0\%$ with recovery beam placement behind the positron beam, and $\eta = 73.5\%$ with the recovery beam in front. 

Since we use the same optimization methods as in the plasma column case, we acknowledge the potential for slightly higher efficiencies than those identified in our study. Further simulations are required to establish the uniformity of $E_z$ and stability of the wake in this regime.

\section{Collider Concepts}
The simulations performed in this study show that it is possible to extract an appreciable fraction of the drive beam's energy from the wake by accelerating both a positron bunch and an electron recovery bunch. This process is repeated over many stages to achieve the desired positron beam energy for collisions~\cite{Lindstrm2021}. The energy stored by the electron recovery bunches must also be utilized to minimize the power consumption of the facility. 

One option for utilizing the energy in the electron recovery bunches is to pass the recovery bunches through an Energy Recovery Linac (ERL)~\cite{Bartnik2020}. The ERL serves as both a decelerator of the recovery bunches and an accelerator of new electron drive bunches. This scheme is likely feasible with existing technology, but might not be viable for two reasons. First, an elaborate beamline is required to separate the drive electron bunch from the trailing positron and electron recovery bunches at the end of a plasma stage. 
The large number of return beamlines, along with associated vacuum chambers and magnets, will increase the cost of the collider facility. The second challenge with the ERL approach is that even if the recovery bunches can be extracted and their energy used to accelerate new drive bunches, there is still the issue of creating new recovery bunches, which limits the overall efficiency of the scheme. 

As an alternative, we may consider the trailing electron bunch as a colliding bunch rather than a recovery bunch. In this scenario, both the electron and positron bunches will be delivered to the IP, albeit with somewhat different beam parameters. For very high energy collisions, the electron and positron bunches should have the same energy (note that this is not necessarily the case for lower energy Higgs Factories~\cite{Foster2023}), but it is possible that other beam parameters such as charge, bunch length and emittance might differ between the two beams. A detailed study using beam-beam codes such as GUINEA-PIG or WarpX is required to better understand both the efficiency and the luminosity-per-power optimization in this scenario.

\begin{figure*}[htbp]
    \includegraphics[width = \linewidth]{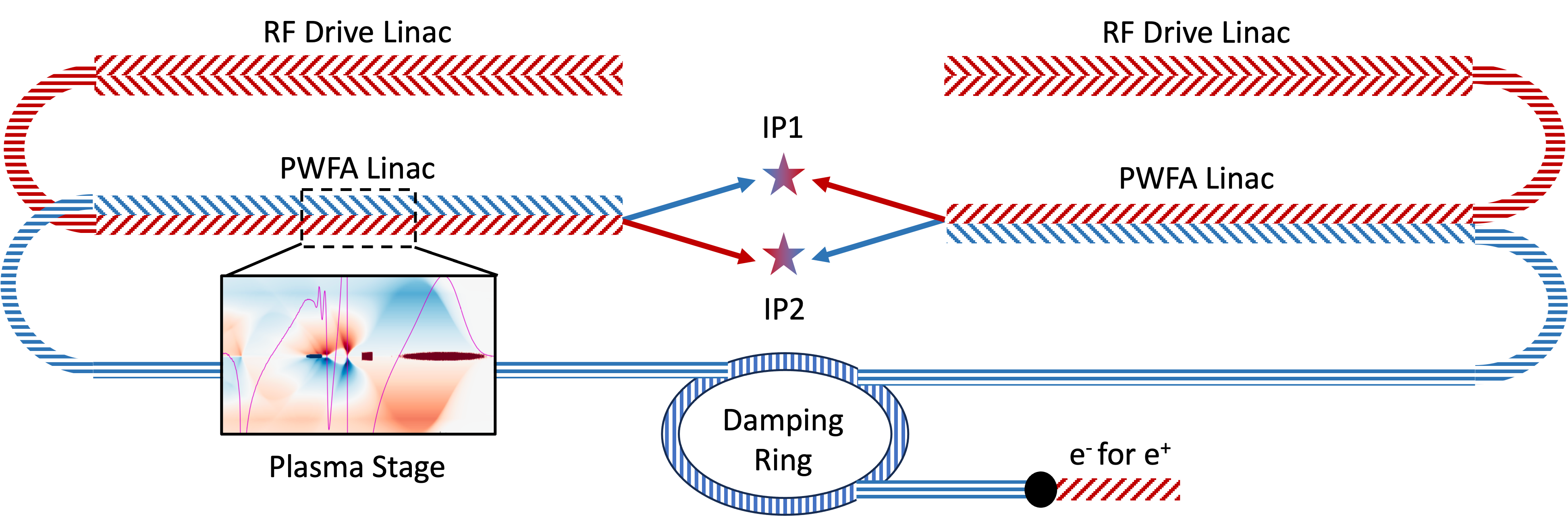}
    \caption{Schematic of a collider with two IPs, with each linac arm of the collider accelerating both electron and positron bunches for collisions. Note that the electron and positron bunches will not have the same properties at the collision point (e.g. charge, bunch length, emittance) and that further studies with beam-beam simulation codes are needed to assess the luminosity-per-power of such a scheme.}\label{fig:fork}
\end{figure*}

Once the beams reach their final energy after many stages of acceleration, they may be delivered to a single IP using an SLC-like configuration~\cite{Richter1980}. To preserve the beam emittance, the size of the arcs should scale as $E^2$, which is an unfavorable scaling for multi-TeV collision energies. A dual-IP collider is considered in Figure~\ref{fig:fork}. Previous collider designs, such as the NLC, have also considered dual IPs~\cite{Raubenheimer2018}. Dual IP designs enable multiple general-purpose detectors to be operated simultaneously, which allows for competition and validation of high-profile particle physics measurements.

\section{Conclusion}
Electron filament PWFA shows promise towards high-quality positron acceleration for a future plasma collider. The realization of such a collider is contingent upon minimizing its environmental impact~\cite{Roser2023,Roser2022} and is challenged by energy efficiency limitations.
Our simulations indicate that the introduction of a secondary electron beam into an electron filament scheme results in a net gain in energy-transfer efficiency. In some cases, specially tailored beam profiles for the trailing beams allowed the average accelerating field to be flattened over the beams. However, the sensitivities experienced with $E_z$ implies that beam quality preservation remains a challenge. Developing a theory on recoverable energy in the blowout regime requires further research.
Moreover, additional simulations are necessary to better understand beam preservation in the energy recovery schemes described in this paper.
Lastly, future studies will examine the stability of these energy recovery schemes by subjecting them to beam offsets at finer resolution.

\section{Data Availability}
Input scripts and analysis can be found at \\
\url{https://github.com/MaxVarverakis/PositronPWFA}.

\section*{Acknowledgments}
Work supported by the U.S. Department of Energy, United States under Contracts DE-AC02-76SF00515 and DE-AC02–05CH11231, and the National Science Foundation, United States (Grants No. PHY-1535696 and No. PHY-2012549).

\appendix
\section{Relevant definitions and notation}\label{appendix}
For the purposes of this paper, we define the following variables and constants: $\xi = z-ct$ is the co-moving coordinate in the speed-of-light frame, $c$ being the speed of light in vacuum, $k_p^{-1} = c/\omega_p$ the plasma skin depth, and $\omega_p = \sqrt{n_0e^2/m\varepsilon_0}$ the plasma frequency for $n_0$ plasma electron density, $e$ the electron charge, $m$ electron mass, and $\varepsilon_0$ vacuum permittivity. Electric and magnetic fields are normalized to the cold, nonrelativistic wave-breaking field $E_0 = \omega_pmc/e$. 

We define the energy-transfer efficiency $\eta$ as the ratio between the energy change of the beams~\cite{Hue2021}, 
\begin{equation}
    \eta = -\left(\frac{Q_p\langle E_z\rangle_p + Q_r\langle E_z\rangle_r}{Q_d\langle E_z\rangle_d}\right),
\end{equation}
where $Q_{p,r,d}$ is the corresponding positron, electron recovery, and drive beam charge and $\langle E_z\rangle_{p,r,d}$ the longitudinal electric field averaged over the beam profiles. In practice, we compute each $Q\langle E_z\rangle$ term as the dot product of the longitudinal electric field with the per-slice charge of the beam. We ignore off-axis contributions to efficiency due to the cylindrical symmetry of the fields in the blowout regime.
\setcounter{section}{1}

\section*{References}
\bibliographystyle{iopart-num}
\bibliography{main}

\end{document}